\documentclass[%
reprint,
superscriptaddress,
amsmath,amssymb,
aps,
pra,
]{revtex4-2}
\usepackage[utf8]{inputenc}
\usepackage[english]{babel}
\usepackage{blindtext}
\usepackage{amsmath}
\usepackage{graphicx}   
\graphicspath{{Figures/}}   
\usepackage{dcolumn} 
\usepackage{bm} 
\usepackage[hidelinks]{hyperref} 
\usepackage[separate-uncertainty=true]{siunitx}  
\usepackage{svg}
\usepackage{lipsum}  

\usepackage[all]{hypcap}
\usepackage[capitalise,nameinlink]{cleveref}
\newcommand{\crefadd}[2]{\hyperref[#1]{\cref{#1}#2}}  
\newcommand{\Crefadd}[2]{\hyperref[#1]{\Cref{#1}#2}}        
\usepackage{xcolor}  
\definecolor{bluegray}{RGB}{40,180,160}
\definecolor{navygray}{RGB}{110,140,170}
\hypersetup{
citecolor={bluegray},
linkcolor={navygray},
urlcolor={bluegray},
}
\usepackage{textcomp}

\usepackage{epstopdf}
\usepackage{braket}
\usepackage{upgreek}
\usepackage{changes}
\usepackage{natbib}
\usepackage[title]{appendix}
\usepackage{physics}

\usepackage{multirow}
\usepackage{booktabs}

\definecolor{color_a}{HTML}{F7403A}
\definecolor{color_b}{HTML}{00814F}
\definecolor{color_c}{HTML}{FFCE00}
\definecolor{color_d}{HTML}{C04191}
\definecolor{color_e}{HTML}{F3A4BA}
\definecolor{color_f}{HTML}{8D5E2A}
\definecolor{color_g}{HTML}{D5C900}
\definecolor{color_h}{HTML}{662483}

\begin{document}

\title{Simultaneous sweet-spot locking of gradiometric fluxonium qubits}

\author{Denis~B\'en\^atre}
\email{denis.benatre@kit.edu}
\affiliation{IQMT,~Karlsruhe~Institute~of~Technology,~76131~Karlsruhe,~Germany}

\author{Mathieu~F\'echant}
\affiliation{IQMT,~Karlsruhe~Institute~of~Technology,~76131~Karlsruhe,~Germany}

\author{Nicolas~Zapata}
\affiliation{IQMT,~Karlsruhe~Institute~of~Technology,~76131~Karlsruhe,~Germany}

\author{Nicolas~Gosling}
\affiliation{IQMT,~Karlsruhe~Institute~of~Technology,~76131~Karlsruhe,~Germany}

\author{Patrick~Paluch}
\affiliation{IQMT,~Karlsruhe~Institute~of~Technology,~76131~Karlsruhe,~Germany}

\author{Thomas~Reisinger}
\affiliation{IQMT,~Karlsruhe~Institute~of~Technology,~76131~Karlsruhe,~Germany}

\author{Ioan~M.~Pop}
\email{ioan.pop@kit.edu}
\affiliation{IQMT,~Karlsruhe~Institute~of~Technology,~76131~Karlsruhe,~Germany}
\affiliation{PHI,~Karlsruhe~Institute~of~Technology,~76131~Karlsruhe,~Germany}
\affiliation{Physics~Institute~1,~Stuttgart~University,~70569~Stuttgart,~Germany}

\date{\today}

\begin{abstract}
Efforts to scale up superconducting processors that employ flux-qubits face numerous challenges, among which is the crosstalk created by neighboring flux lines, which are necessary to bias the qubits at the zero-field and $\Phi_0/2$ sweet spots.
A solution to this problem is to use symmetric gradiometric loops, which incorporate a flux locking mechanism that, once a fluxon is trapped during cooldown, holds the device at the sweet spot and limits the need for active biasing.
We demonstrate this technique by simultaneously locking multiple gradiometric fluxonium qubits in which an aluminum loop retains the trapped fluxon indefinitely.
By compensating the inductive asymmetry between the two loops of the design, we are able to lock the effective flux-bias within $\Phi_\mathrm{eff} = \SI{-3e-4}{} \,\Phi_0$ from the target, corresponding to only $\SI{15}{\percent}$ degradation in $T_{2,\mathrm{E}}$ when operated in zero external field.
The design strategy demonstrated here reduces integration complexity for flux qubits by minimizing cross-talk and potentially eliminating the need for local flux bias.
\end{abstract}


\maketitle 


The last decade has seen rapid developments in flux-qubit engineering and design, parallel to those in transmons.
In this type of superconducting qubits, the states are encoded in symmetric and antisymmetric superpositions of counterpropagating current states in an inductive superconducting loop, which includes at least one Josephson junction to give its non-linearity.  
Flux qubits generally have higher anharmonicity compared to transmons, which allows faster gates, and are insensitive to charge offset thanks to their inductive shunt \cite{koch_charging_2009}. Using superinductors \cite{manucharyan_fluxonium_2009, nguyen_high-coherence_2019,zhang_universal_2021}, the so-called fluxonium qubit can be designed to reach state-of-the art coherence times above $\SI{1}{\milli s}$ \cite{somoroff_millisecond_2023} similarly to the most recent improvements in transmons \cite{bland_2d_2025}.

While quantum non-demolition readout \cite{gusenkova_quantum_2021} and high-fidelity single-qubit and two-qubit gates have been demonstrated with fluxoniums \cite{xiong_arbitrary_2022, bao_fluxonium_2022, moskalenko_high_2022, ding_high-fidelity_2023, rower_suppressing_2024}, implementations of multi-fluxonium processor designs remain a rarity. Scaling up fluxonium processors has its specific challenges \cite{nguyen_blueprint_2022}, among which is the crosstalk created by neighboring flux lines. 
Characterizing and inverting the flux-crosstalk matrix becomes increasingly computationally demanding with the number of qubits, couplers and resonators on chip and is rendered difficult by the non-linearity of the joint frequency response of these elements \cite{dai_calibration_2021, barrett_learning-based_2023}. Additionally, it requires frequent recalibration due to flux offset drifts and jumps over a cooldown.
One solution could be 'pre-biased' flux qubits, such as the ones realized with so-called $\pi$-junctions made from ferromagnetic materials \cite{kontos_josephson_2002, kim_superconducting_2024}, however they complicate fabrication and may introduce new decoherence channels.

\begin{figure}[tp!]
    \centering
    \includegraphics[width=1\linewidth]{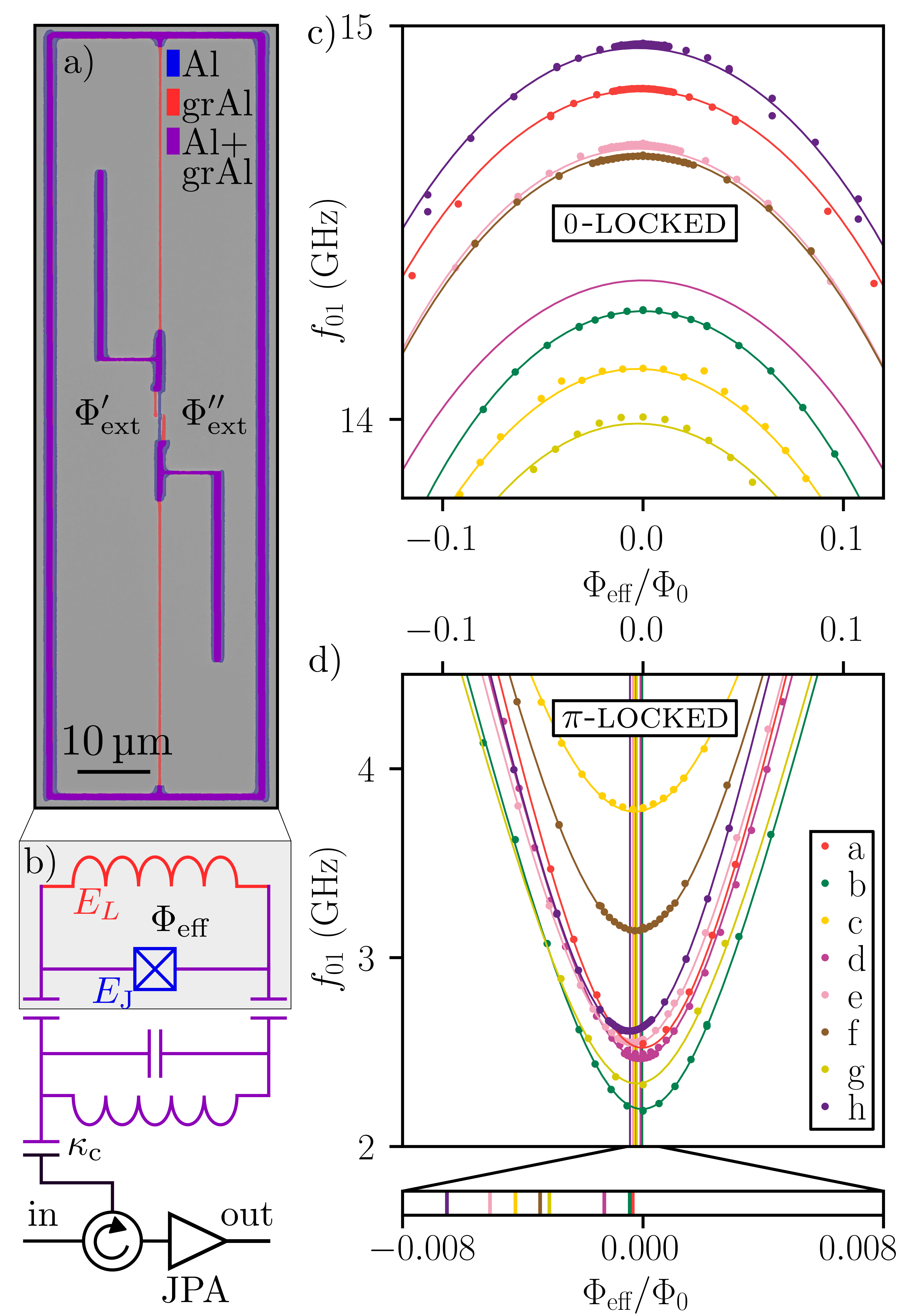}
    \caption{Demonstration of flux locking. \textbf{(a)} False-colored microscope picture of the gradiometric fluxonium featuring a standard Al/$\mathrm{AlO}_x$/Al junction (blue), Al capacitive pads to couple to the resonator (purple) and granular aluminum (grAl) inductors with normal-state resistivity $\rho_n \approx \SI{2e3}{\micro\ohm \cm}$ and thickness $\SI{40}{nm}$ (red). Purple parts are stacks of grAl above two layers of Al shunt. The fluxes threading the two loops $\Phi'_\mathrm{ext}$ and $\Phi''_\mathrm{ext}$ give rise to an effective flux bias $\Phi_\mathrm{eff}$ (see \cref{eq:Phi_eff}). \textbf{(b)} Reflection measurement setup. All samples, each consisting of a qubit capacitively coupled to a LC resonator, are placed in a single 3D waveguide and measured in reflection by a Josephson Parametric Amplifier (JPA) \cite{winkel_nondegenerate_2020}. \textbf{(c)} Spectra of 8 gradiometric fluxonium samples measured in the vicinity of $\Phi_\mathrm{eff}=0$, after cooling down either in zero field or \textbf{(d)} in a field ${\Phi_0}/{2} < \Phi'_\mathrm{ext}+\Phi''_\mathrm{ext} < {3\Phi_0}/{2}$,  corresponding to $\Phi_0$ trapped in the aluminum ring of the device. Markers depict the frequency of the $0-1$ transition extracted from a two-tone spectroscopy and lines represent a fit to the fluxonium Hamiltonian in \cref{eq:hamiltonian}. Vertical lines indicate the position of the minimal frequency and are shown in detail in the bottom inset.}
    \label{fig:fig1}
\end{figure}

In this work, we propose to exploit the flux-locking mechanism arising from flux quantization in gradiometric flux-qubits \cite{schwarz_gradiometric_2013, gusenkova_operating_2022}.
Cooling a superconducting ring in a magnetic field $B_\mathrm{cd}$, then removing it at base temperature, leaves a persistent current corresponding to a magnetic flux value given by the nearest integer multiple of $\Phi_0$ \cite{deaver_experimental_1961, doll_experimental_1961}.
The gradiometric fluxonium (see \crefadd{fig:fig1}{a}) is built by splitting the ring into two inductive loops connected by a Josephson junction. Following Ref.~\cite{gusenkova_operating_2022}, this type of fluxonium is modeled by the Hamiltonian 
\begin{equation}
    \label{eq:hamiltonian}
    \hat{\mathcal{H}} = 4E_C \hat{n}^2 + \frac{E_L}{2} \! \left( \hat{\phi} +  \varphi_\mathrm{eff} \!+\! 2\pi m \frac{1+\alpha}{2} \! \right)^2 \! - \! E_\mathrm{J} \cos\hat{\phi},
\end{equation}
where $m$ is the number of trapped fluxons corresponding to $B_\mathrm{cd}$ and the area of the ring, $\alpha=(L''-L')/(L''+L')$ is the inductive asymmetry between the loops, and the effective flux bias induced by a uniform applied field $B_\mathrm{ext}$ is given by
\begin{equation}
    \label{eq:Phi_eff}
    \Phi_\mathrm{eff} = \varphi_\mathrm{eff} \frac{\Phi_0}{2\pi}  = \frac{\Phi'_\mathrm{ext} - \Phi''_\mathrm{ext}}{2} + \alpha \frac{\Phi'_\mathrm{ext} + \Phi''_\mathrm{ext}}{2},
\end{equation}
where $\Phi'_{\mathrm{ext}}$ and $\Phi''_{\mathrm{ext}}$ are the external fluxes created by $B_\mathrm{ext}$ in each loop. 
Note that the effective flux $\Phi_\mathrm{eff}$ only designates the flux created by the applied magnetic field $B_\mathrm{ext}$, which we distinguish from the flux arising from the trapped fluxons and given by $m (1+\alpha)/2$, and that magnetic field gradients increase the effective flux bias via the first term in \cref{eq:Phi_eff}.

For even numbers of fluxons $m$ trapped in the ring, including $m=0$, the fluxonium is locked at the zero flux sweet spot (hereafter designated as $0$-locked). For odd $m$, the qubit is locked at the half-flux-quantum sweet spot, $\pi$-phase biasing the junction ($\pi$-locked).  
Following \cref{eq:Phi_eff}, in the ideal case of a completely symmetric sample, the qubit is only susceptible to field gradients, suppressing crosstalk from homogeneous fields and removing the need for static field bias, which additionally reduces the heat load in the cryostat.
As the magnetic field is proportional to $1/d^2$, with $d$ the distance to the next neighbor's flux line, the gradient is proportional to $1/d^3$ which reduces flux crosstalk by orders of magnitude (see Appendix \ref{subsec:flux_crostalk}).
In a real device, some asymmetry is unavoidable. By tuning the area and inductive asymmetries of the loops, both the effective area $A_\mathrm{eff}= \Phi_\mathrm{eff}/B_\mathrm{ext} $ and the remaining flux offset at zero field $\text{sgn}(A_\mathrm{eff})m\alpha/2$ can in principle be made arbitrarily small, independently.

The samples (cf.~\crefadd{fig:fig1}{a}) are fabricated in a single e-beam lithography step, using a three-angle shadow evaporation, similarly to Ref.~\cite{grunhaupt_granular_2019} (cf.~\cref{subsec:fabrication}). Six chips, each hosting one to two qubits, are mounted in the same 3D copper waveguide, equipped with a global magnetic field coil, similar to Refs.~\cite{Kou2018Jun, winkel_implementation_2020}. We estimate the relative contribution to $\Phi_\mathrm{eff}$ from magnetic field gradients to be $10^{-4}$ via the first term in \cref{eq:Phi_eff}, which in the following can be neglected.
We perform standard dispersive microwave reflection measurements at $\SI{30}{mK}$, as shown in \crefadd{fig:fig1}{b}. In order to achieve single-shot readout, we employ a Josephson parametric amplifier (JPA) \cite{winkel_nondegenerate_2020} operated in reflection.

In \cref{fig:fig1} we show a demonstration of the flux-locking mechanism at the zero flux and at the half-flux-quantum sweet spots for qubits a--h, as visible in the measured $0-1$ transition frequency for $m=0$ (\crefadd{fig:fig1}{c}) and $m=1$ (\crefadd{fig:fig1}{d}) fluxons trapped in the ring. 
We can achieve $m=1$ by cooling down in a magnetic field $B_\mathrm{cd} \approx \SI{0.66}{\micro T}$ corresponding to $\SI{1}{\Phi_0}$ in the ring area of $A_\mathrm{ring}=107\,\times\,\SI{29.4}{\micro m^2}$.
The $0$-locked and $\pi$-locked data are fitted together to a fluxonium spectrum (cf.~\cref{eq:hamiltonian}) from which we extract the Josephson energy $E_\mathrm{J}$, the capacitive energy $E_C$, the inductive energy $E_L$ (see \cref{tab:samples} and \cref{subsec:spectra}) and the effective area $A_\mathrm{eff}$.
Spectra vs. magnetic field are shown in \Cref{subsec:spectra}, illustrating the variation of $A_\mathrm{eff}$.
All samples have similar fluxonium parameters, with mean values $\bar{E}_\mathrm{J}/h=\SI{8.8}{GHz}$, $\bar{C}=\SI{4.7}{fF}$ (with about $\SI{2.5}{fF}$ coming from the intrinsic capacitance of the junction) and $\bar{L}=\SI{88}{nH}$, corresponding to a sheet inductance $\bar{L}_\mathrm{\square, grAl}=\SI{0.22}{nH}/\square$.

\begin{table}[tp]
    \centering
    \begin{tabular}{cccc}
        \toprule
        Sample & $E_\mathrm{J}/h$ (GHz) & $E_C/h$ (GHz) & $E_L/h$ (GHz) \\
        \midrule
        $\textcolor{color_a}{\bullet}$ a & $\num{9.21(0.16)}$ & $3.97 \pm 0.07$ & $1.95 \pm 0.05$ \\ 
        $\textcolor{color_b}{\bullet}$ b & $\num{9.19(0.16)}$ & $4.05 \pm 0.04$ & $1.59 \pm 0.04$ \\ 
        $\textcolor{color_c}{\bullet}$ c & $\num{7.50(0.13)}$ & $4.48 \pm 0.07$ & $1.96 \pm 0.03$ \\ 
        $\textcolor{color_d}{\bullet}$ d & $\num{8.87(0.23)}$ & $3.88 \pm 0.03$ & $1.87 \pm 0.03$ \\ 
        $\textcolor{color_e}{\bullet}$ e & $\num{9.06(0.15)}$ & $4.09 \pm 0.06$ & $1.84 \pm 0.05$ \\ 
        $\textcolor{color_f}{\bullet}$ f & $\num{8.45(0.15)}$ & $4.16 \pm 0.05$ & $2.05 \pm 0.05$ \\ 
        $\textcolor{color_g}{\bullet}$ g & $\num{8.90(0.14)}$ & $4.09 \pm 0.08$ & $1.58 \pm 0.04$ \\ 
        $\textcolor{color_h}{\bullet}$ h & $\num{9.22(0.15)}$ & $3.97 \pm 0.07$ & $2.01 \pm 0.06$ \\
        \bottomrule
    \end{tabular}
    \caption{Qubit parameters for samples a--h}
    \label{tab:samples}
\end{table}

A previous implementation \cite{gusenkova_operating_2022} with a granular aluminum ring has shown flux escape at base temperature, mediated by ionizing radiation interactions with the substrate in the vicinity of the device. In above ground conditions these flux escapes occurred on a timescale of one hour, and they were completely removed when operating in an underground laboratory shielded from ionizing radiation.
We solve this issue by using a wide, pure aluminum ring.
Due to its orders-of-magnitude higher superfluid stiffness ($\rho_\mathrm{s,Al}/\rho_\mathrm{s,grAl} \sim 100 $) 
persistent currents in aluminum are less susceptible to phase slips.
As a result, we never observe flux escapes after locking (cf.~\Cref{subsec:flux_escape} for a time trace over 50 hours).
However, we typically observe simultaneous $\pi$-locking for only 80\% of the qubits, which means that in each cooldown one or two devices lock to even values of $m$ instead of odd. This behavior is unexpected and could be due to ionizing radiation activated flux escapes happening during the cooldown, when the superfluid stiffness $\rho_\mathrm{s,Al}$ is suppressed close to the superconducting critical temperature, rendering the ring vulnerable to phase slips.
Potential mitigation strategies align with those used to address quasiparticle poisoning in superconducting circuits, such as employing quasiparticle traps adjacent to the sample \cite{riwar_normal-metal_2016} or shielding from cosmic radiation underground \cite{gusenkova_operating_2022}. An additional strategy could involve accelerating the cooldown or enhancing sample thermalization to minimize the time spent near the transition.

\begin{figure}
    \centering
    \includegraphics[width=1\linewidth]{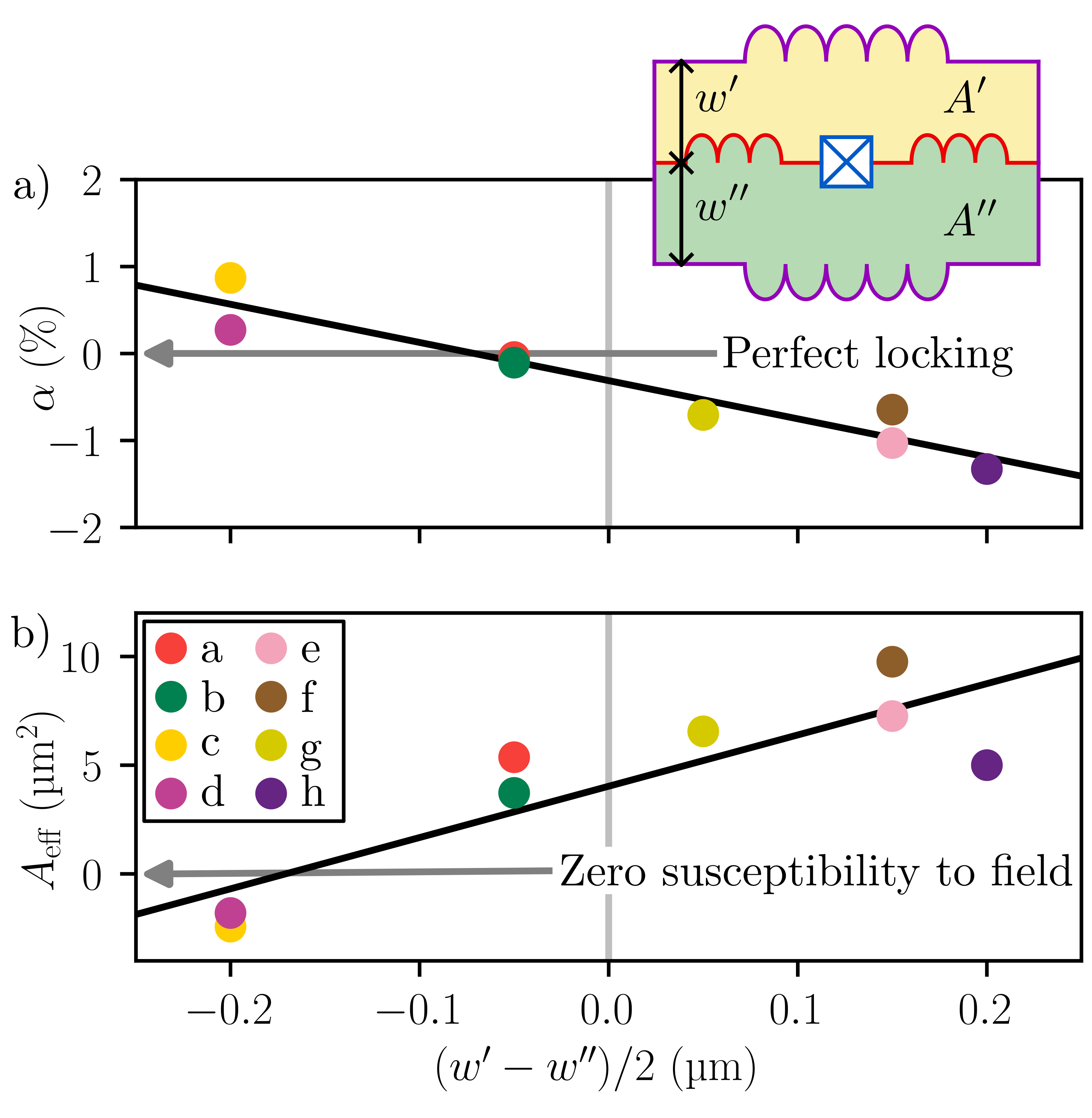}
    \caption{Tuning of the asymmetry. Inset: Circuit diagram of the gradiometric fluxonium, showing the Josephson junction (blue), the granular aluminum inductances (red) in the middle, and the low-value inductances (purple) of the aluminum ring. The difference in loop widths $w'-w''$ is tuned while keeping the total size constant.  \textbf{(a)} Inductive asymmetry  $\alpha=(L''-L')/(L''+L')$ vs. loop width asymmetry. $\alpha$ is obtained from the offset of the spectrum when the fluxonium is $\pi$-locked (cf.~\crefadd{fig:fig1}{d}), such that $\Phi_\mathrm{offset}/\Phi_0=\alpha/2$.
    Perfect sweet-spot locking is realized when $\alpha=0$.
    \textbf{(b)} Effective area $A_\mathrm{eff}$ vs. loop width asymmetry. $A_\mathrm{eff}$ is extracted from fitting the spectra together with fluxonium parameters $E_\mathrm{J}$, $E_C$ and $E_L$. Zero susceptibility to homogeneous magnetic field is achieved when $A_\mathrm{eff}=0$.}
    \label{fig:fig2}
\end{figure}

The inset of \crefadd{fig:fig1}{d} displays the measured flux offsets from perfect sweet-spot locking across the samples, resulting from a deliberate sweep of the loop width asymmetry $(w'-w'')/2$ (inset of \crefadd{fig:fig2}{a}) over the range $\SI{-200}{nm}$ to $\SI{200}{nm}$. 
We observe a linear dependence of $\alpha$ (\crefadd{fig:fig2}{a}) and $A_\mathrm{eff}$ (\crefadd{fig:fig2}{b}) on the asymmetry, as expected. 
From the slopes, we extract a surface change of $\SI{93}{\micro m ^2 / \micro m }$ and a change in $\alpha$ corresponding to $\SI{-0.12}{nH/\micro m}$, assuming the aluminum ring sheet inductance $L_\square=\SI{10}{pH}/\square$ (cf.~\Cref{subsec:asymmetry}).
While the surface change aligns with the device’s $\SI{107}{\micro m}$ length, the observed $\alpha$ change is roughly three times larger than expected for a constant sheet inductance around the loop.
This discrepancy suggests a larger than expected inductive contribution of the horizontal wires in \crefadd{fig:fig1}{a}.
The zero crossings of $\alpha$ and $A_\mathrm{eff}$ occur at different values of $(w'-w'')/2$, shifted away from zero.
This observation implies that adjusting $w'-w''$ alone cannot offer a device that simultaneously locks perfectly to the sweet spots and remains insensitive to magnetic fields.
The shift in $A_\mathrm{eff}$ corresponds to a $\SI{100}{n m }$ geometric offset, which is explained by the persistent current path through the two aluminum layers with different kinetic inductances (cf. Appendix \ref{subsec:area_asymmetry}).
The shift in $\alpha$ is explained by a small difference in the inductance of the left and right loops of $\SI{20}{pH}$, whose origin could be a combination of the geometric inductance and the supercurrent path factors mentioned above.

\begin{figure}
    \centering
    \includegraphics[width=1\linewidth]{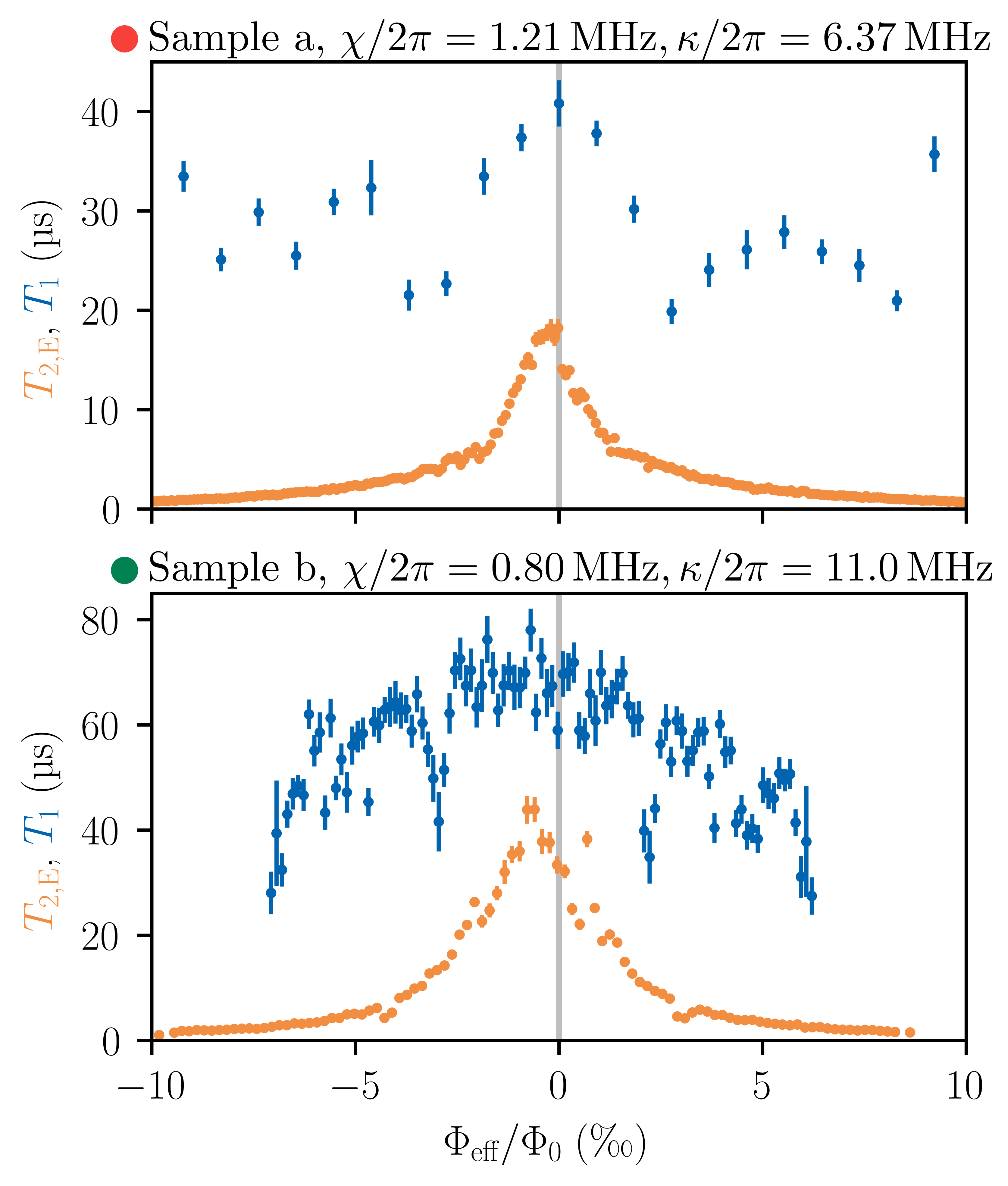}
    \caption{Qubit time-domain characterization. $T_1$ (in blue) and $T_{2,\mathrm{E}}$ (in orange) coherence times for samples $\textcolor{color_a}{\bullet}$ a (top) and $\textcolor{color_b}{\bullet}$ b (bottom) with $m=1$ fluxon trapped vs. effective flux $\Phi_\mathrm{eff}$ around zero field, illustrating the precision of the locking mechanism. Markers show the measured values with uncertainty. The gray vertical line indicates zero field. In each title, $\chi$ is the dispersive shift at the sweet spot and $\kappa$ is the resonator's linewidth.}
    \label{fig:fig3}
\end{figure}

To investigate the locking accuracy in more detail, we measure $T_{2,\mathrm{E}}$ around zero field for samples \textit{a} and \textit{b}, which are the closest to zero in \crefadd{fig:fig2}{a}. In \cref{fig:fig3}, we report $T_\mathrm{2,E}$ values of respectively $\SI{20}{\micro s}$ and $\SI{44}{\micro s}$ at $\Phi_\mathrm{eff}=\SI{-3e-4}{}\,\Phi_0$ and $\Phi_\mathrm{eff}=\SI{-5e-4}{}\,\Phi_0$, similar to values reported for granular aluminum fluxoniums in Ref.~\cite{grunhaupt_granular_2019}. At zero field, they correspond to degradations of $\SI{15}{\percent}$ and $\SI{32}{\percent}$ respectively.
For both samples, $T_{2,\mathrm{E}}$ at the sweet spot follows an exponential decay, not limited by $T_1$, which could be explained, for example, by critical current noise or photon shot noise with $n_\mathrm{res}\approx 0.01$ residual photons in the resonator.
Additionally, \Cref{subsec:coherence_over_time} reveals significant correlated fluctuations in $T_1$ and $T_{2,\mathrm{E}}$ over several hours, as commonly observed in the literature \cite{nguyen_high-coherence_2019}, though their origin remains unknown.
The coherence times measured on both devices are larger than previously measured gradiometric fluxoniums with granular aluminum in the outer ring \cite{gusenkova_operating_2022}, which suggests no detrimental effect of the aluminum ring on the coherence.

In conclusion, we demonstrate simultaneous and stable sweet-spot flux locking of gradiometric fluxoniums, enabled by flux quantization in an aluminum ring. These devices exhibit no flux escapes, unlike previously reported gradiometric fluxoniums with granular aluminum rings. We are able to reduce the field susceptibility and asymmetry of our samples down to an effective area of a few $\si{\micro m^2}$ and an inductive asymmetry less than 1\,\textperthousand. The latter implies that in zero field fluxonium qubits can be locked within $\Phi_\mathrm{eff}=\SI{-3e-4}{}\,\Phi_0$ from the half-flux-quantum sweet spot, with $T_{2,\mathrm{E}}$ reduced by only $\SI{15}{\percent}$ from its maximum.
As an outlook, we anticipate this scheme to be beneficial for large scale integration, and the stability of the flux locking to allow the implementation of more complex systems of trapped fluxons in superconducting circuits \cite{petrescu_fluxon-based_2018}.

\section*{Acknowledgements}
During writing of this manuscript, we became aware of similar results obtained in the research group of Y. Nakamura~\cite{hida_flux-trapping_2025}. We are thankful to K. Matsuura, K. Hida and Y. Nakamura for productive discussions and for their feedback on our manuscript. 
We also thank M. Spiecker for discussions during this work. 
We acknowledge technical support from S. Diewald, and L. Radtke. 
Funding was provided by the Horizon Europe program via Project No.~101113946 OpenSuperQPlus100. 
M.F.~acknowledges funding from the German Ministry of Education and Research (BMBF) within project QSolid (FKZ:13N16151).
N.Z.~acknowledges funding from the Deutsche Forschungsgemeinschaft (DFG – German Research Foundation) under project number 450396347 (GeHoldeQED).
N.G., P.P. and T.R.~acknowledge funding from the German Ministry of Education and Research (BMBF) within project GEQCOS (FKZ:~13N15683).
The facilities use was supported by the KIT Nanostructure Service Laboratory (NSL). We acknowledge qKit for providing a convenient measurement software framework.

\appendix

\begin{figure}[ht!]
    \centering
    \includegraphics[width=1\linewidth]{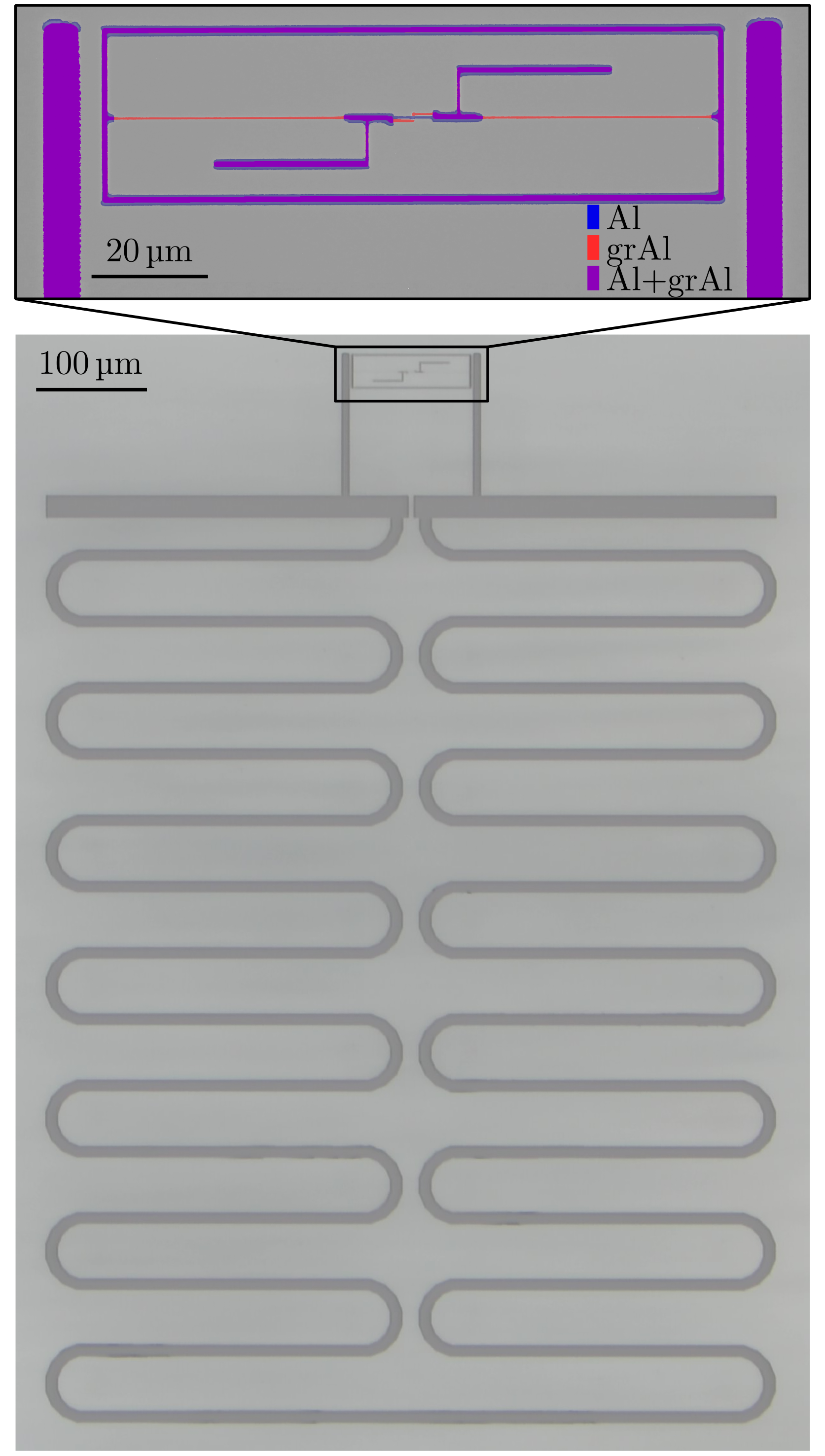}
    \caption{Optical microscope pictures of the qubit (top) and the aluminum LC resonator (bottom) for one of the samples. Each sample's resonator is designed with a different number of meanders in order to spread the frequencies between 6 and 8 $\si{GHz}$. Each chip contains two of such samples and does not have any ground plane.}
    \label{fig:qubit_and_res}
\end{figure}

\begin{figure}[ht]
    \centering
    \includegraphics[width=1\linewidth]{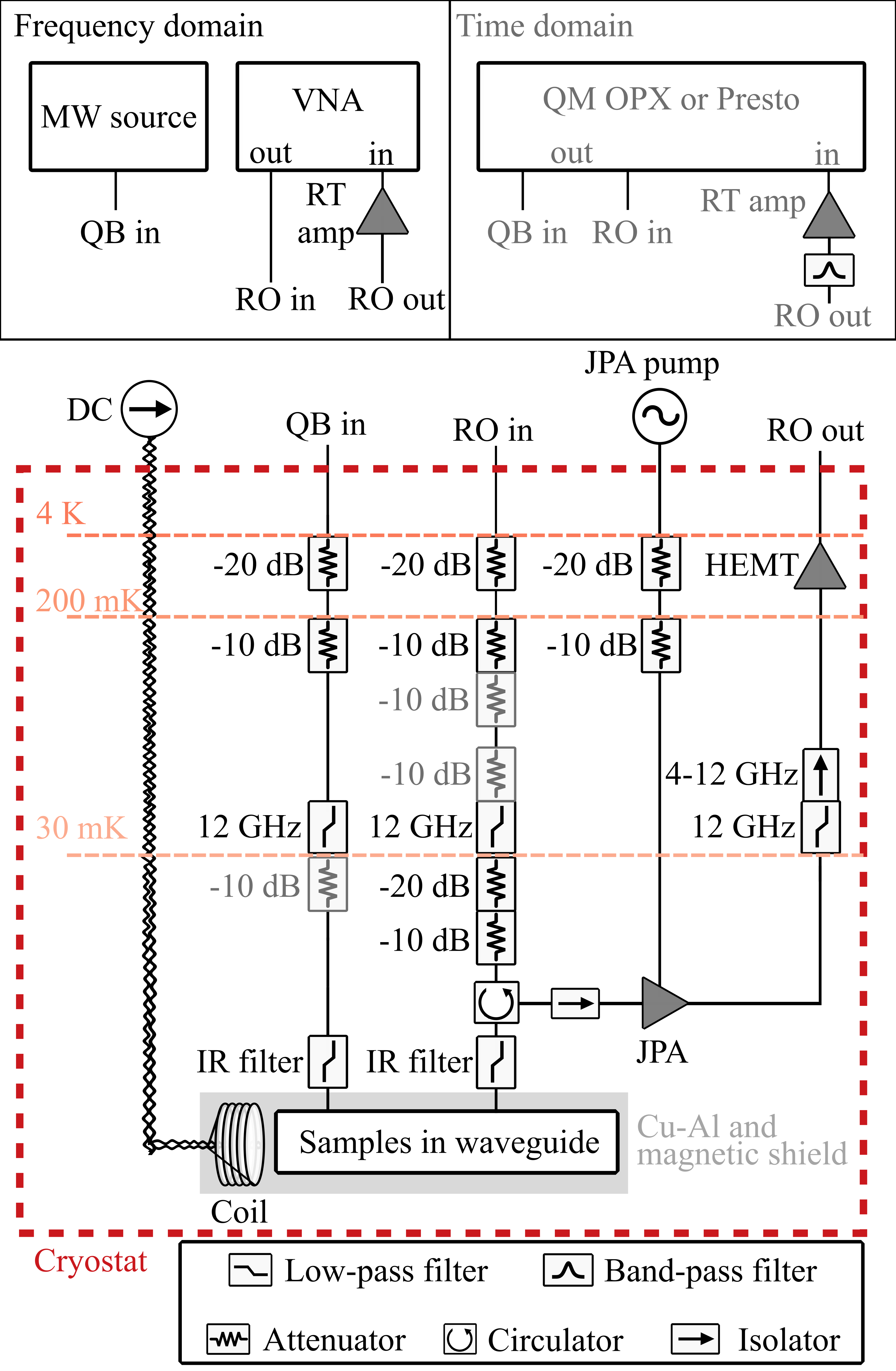}
    \caption{Cryostat and measurement setup. For frequency-domain experiments, we use a Vector Network Analyzer (VNA) and a microwave source as indicated in the top-left corner. The cryostat is set up with all black components, in particular the total attenuation is 60 dB on the readout input line is  and 30 dB on the qubit drive line. For time-domain experiment, we use a Quantum Machines' Operator-X or an Intermodulation Products' Presto. The cryostat is set up with the additional attenuators in grey, increasing the total attenuation to 80 dB on the readout input line and 40 dB on the qubit drive line.}
    \label{fig:cryostat}
\end{figure}

\section{Sample fabrication}
\label{subsec:fabrication}
All samples shown in this paper were fabricated on a single double-side polished c-plane 330-$\si{\micro m}$-thick sapphire wafer. A double stack of MMA(8.5)MAA EL13 and 950 PMMA A4 resists was spincoated on the wafer and patterned using a $\SI{50}{keV}$ e-beam lithography tool. After developing in a IPA/$\mathrm{H}_2$O 3:1 solution for $\SI{2}{min} \, \SI{30}{s}$, aluminum was deposited in a three-angle shadow evaporation process in a Plassys e-beam evaporation system. The thin film deposition recipe includes plasma cleaning and Ti gettering to clean the sample and the chamber, $\SI{-31}{\degree}$-tilted aluminum evaporation at $\SI{1}{nm/s}$ for \SI{20}{s}, 6-$\si{min}$ oxidation at $\SI{11.54}{mbar}$, $\SI{31}{\degree}$-tilted aluminum evaporation at $\SI{1}{nm/s}$ for \SI{30}{s}, argon milling to ensure contact, and granular aluminum deposition consisting of a $\SI{0}{\degree}$ aluminum evaporation at $\SI{1}{nm/s}$ for \SI{36}{s} under a constant oxygen flow at $\SI{5.0}{sccm}$. After lift-off in hot acetone ($\SI{50}{\degree C}$, $\SI{13}{h}$) and hot NEP ($\SI{90}{\degree C}$, $\SI{27}{min}$), the wafer was cleaned in acetone and IPA. Finally, the wafer was diced in $\SI{15}{mm}\,\times\,\SI{3}{mm}$ chips after coating it with a protective S1818 G2 photo-resist, which is removed in hot DMSO ($\SI{80}{\degree C}$, $\SI{10}{min}$). Each chip contains one to two working samples such as the sample shown in \cref{fig:qubit_and_res}.

\begin{figure}[t]
    \centering
    \includegraphics[width=1\linewidth]{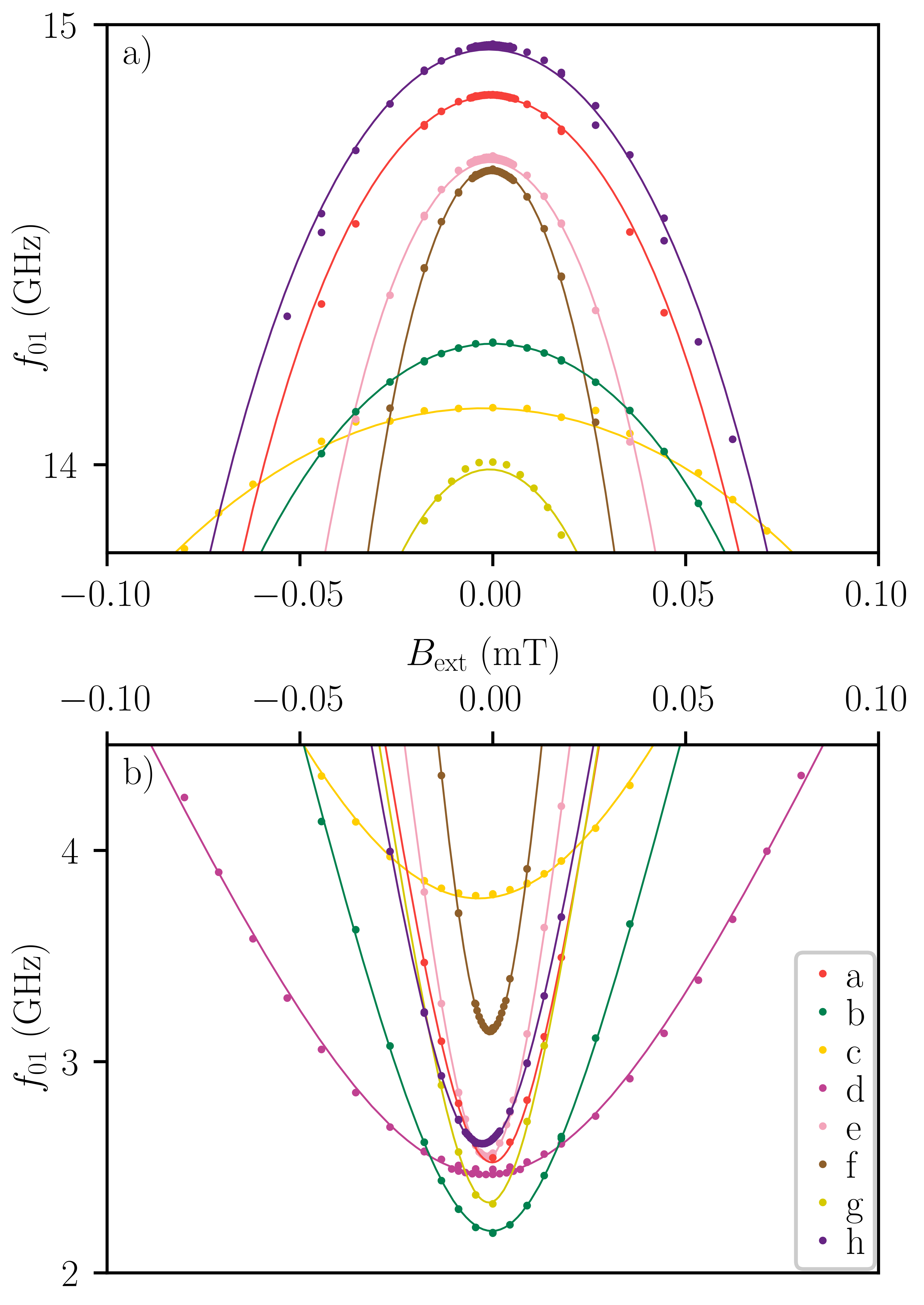}
    \caption{Spectra of all eight samples in the vicinity of zero field, in the $0$-locked (a) and $\pi$-locked (b) cases, as a function of the applied field $B_\mathrm{ext}$, as opposed to the effective flux in \cref{fig:fig1}.}
    \label{fig:spectra_field}
\end{figure}

\begin{figure*}[ht!p]
    \centering
    \includegraphics[width=1\linewidth]{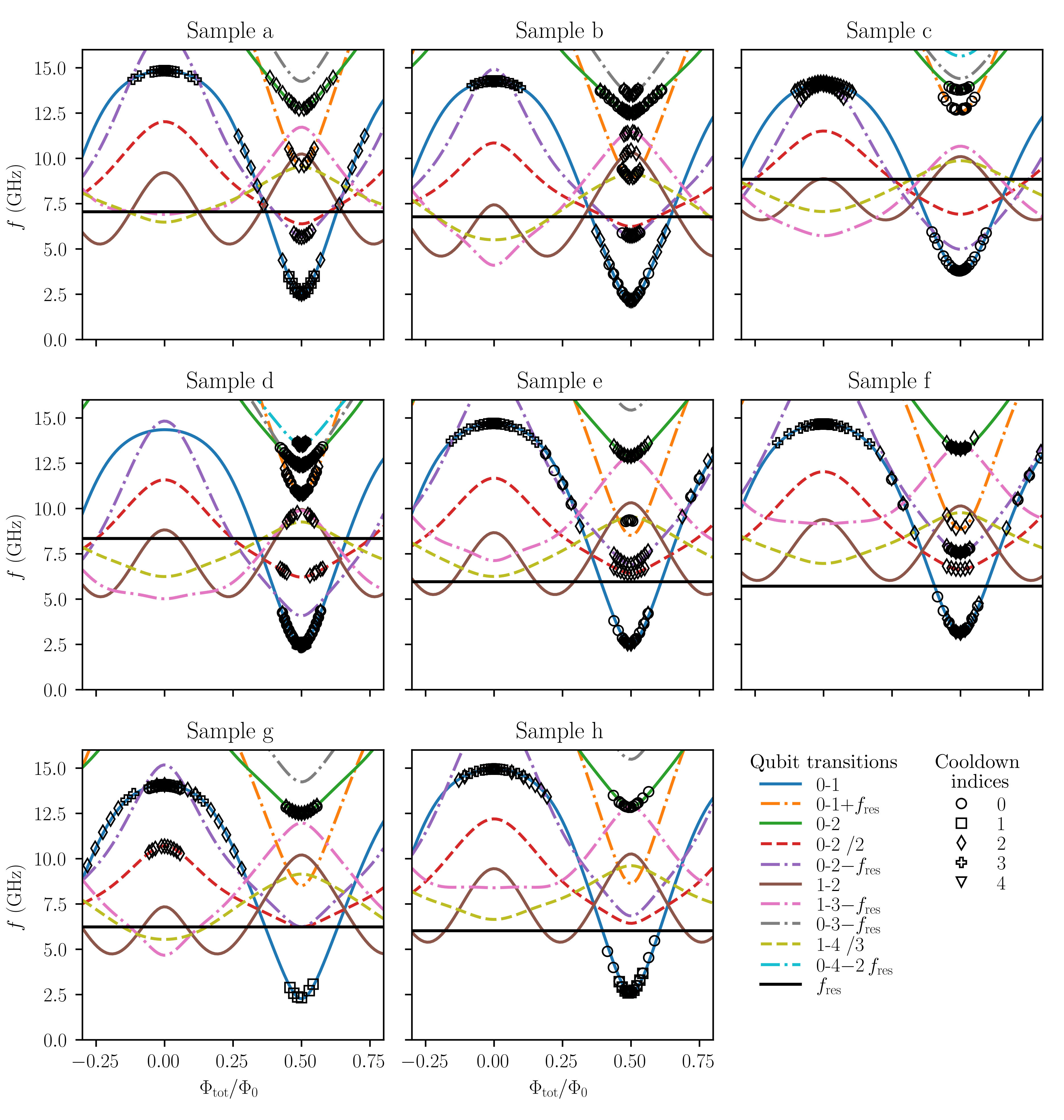}
    \caption{Spectra with fits of all eight samples shown in the main text. Markers depict the transition frequencies extracted from a two-tone spectroscopy, with marker type indicating different cooldowns. Colored solid lines represent the main transition frequencies, dashed lines represent multiple-photon transitions and dash-dotted lines represent transitions mediated by resonator photons. Black solid lines represent the readout resonator frequency $f_\mathrm{res}$ of each sample.}
    \label{fig:spectra_and_fits}
\end{figure*}

\section{Cryostat setup}
\label{subsec:cryostat}

All experiments were performed in a Qinu Sionludi XL cryostat, at a base temperature of 20 to $\SI{30}{mK}$, over different cooldowns, with a setup shown in \cref{fig:cryostat}. Frequency-domain experiments were performed using a Vector Network Analyzer (VNA) to readout the resonator and a microwave source to drive the qubits. The total attenuation was \SI{60}{dB}, spread between the \SI{4}{K}, the \SI{200}{mK} and the \SI{30}{mK} stages, for the readout line and \SI{30}{dB} on the qubit drive line. For these measurements, the JPA was kept turned off and no bandpass filter was used at the output.
For time-domain experiments, the total attenuation on the readout line was increased to \SI{80}{dB} as represented in grey in \Cref{fig:cryostat}, in order to reduce the photon shot noise. The attenuation on the qubit drive line was also increased by adding a \SI{10}{dB} attenuator on the \SI{30}{mK} stage. The JPA was turned on and the room-temperature band-pass filter, used to filter out the strong JPA pump signal, was tuned to the measured resonator frequency. The time-domain experiments were performed using an OPX from Quantum Machines (data in \Cref{fig:fig3}) and a Presto from Intermodulation Products (additional measurements).

\begin{figure}[t!]
    \centering
    \includegraphics[width=1\linewidth]{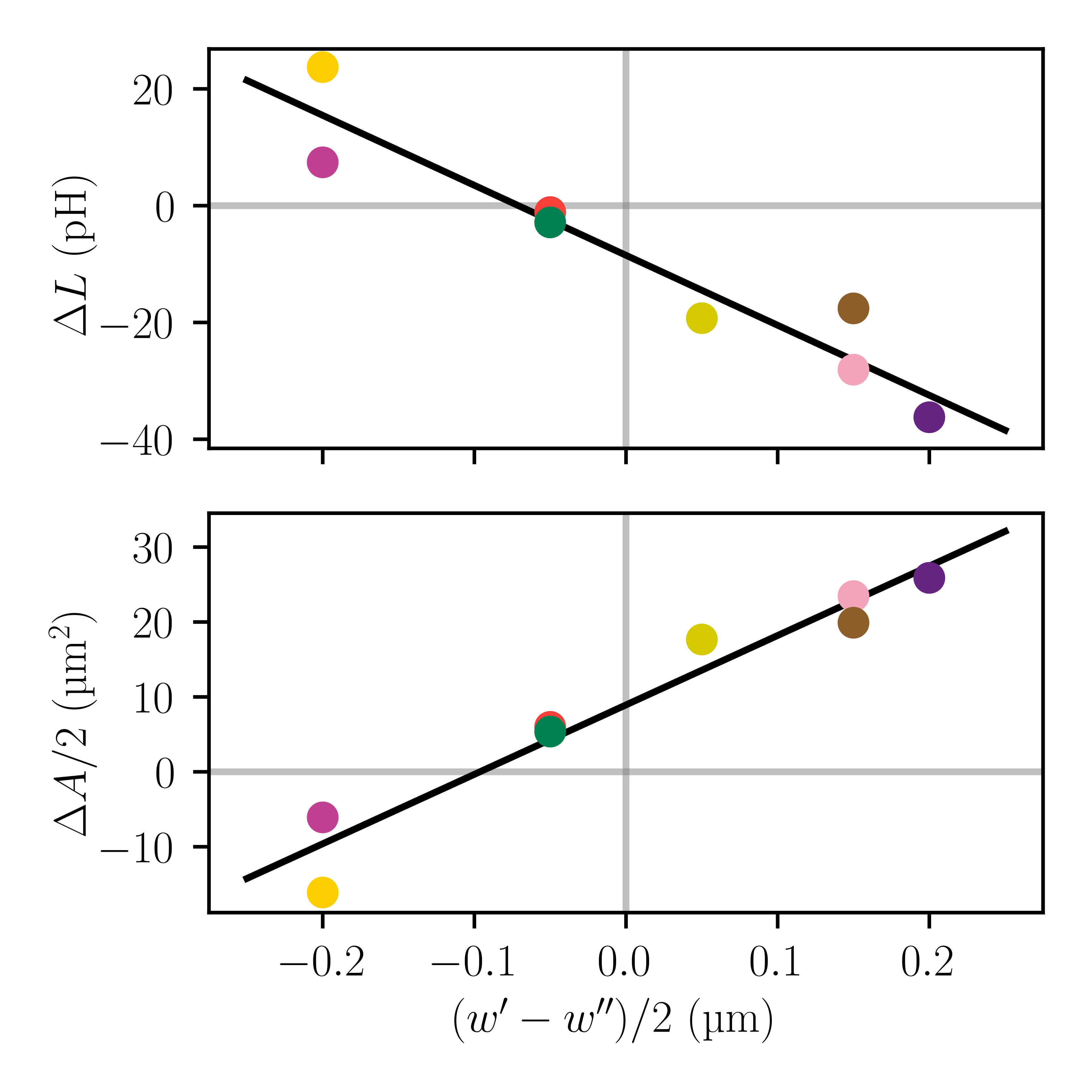}
    \caption{Inductance and area difference between the loops as a function of the asymmetry $(w'-w'')/2$. The inductance is calculated by assuming a sheet inductance of $\SI{10}{pH}/\square$ for the aluminum ring. $\Delta A /2 $ is calculated from the measured $A_\mathrm{eff}$ and $\alpha$ and fits with the expected variation.}
    \label{fig:DeltaL_DeltaA}
\end{figure}

\section{Spectra}
\label{subsec:spectra}

We measured the spectra of the devices shown in the main text by a standard continuous wave two-tone spectroscopy in reflection on the readout resonator of each sample individually. The transition frequencies were then extracted by fitting a Lorentzian function at each field.
To illustrate the different $A_\mathrm{eff}$ measured in our samples, we show in \Cref{fig:spectra_field} the spectra of the same eight samples as in the main text, in the $0$-locked (a) and $\pi$-locked (b) cases, as a function of the applied magnetic field $B_\mathrm{ext}$. 

All eight samples were fitted to the fluxonium Hamiltonian using the scqubits Python package \cite{groszkowski_scqubits_2021}. On top of the fluxonium parameters $E_\mathrm{J}$, $E_C$ and $E_L$, we also fitted the effective area $A_\mathrm{eff}$ that gives the scaling in $\Phi_\mathrm{eff}$. The data used in the fits include the $0-1$ transition in the $0$-locked and $\pi$-locked case, as well as higher transitions such as $0-2$, multi-photon transitions and transitions mediated by resonator photons. The fits are shown in \Cref{fig:spectra_and_fits} as a function of $\Phi_\mathrm{tot} = \Phi_\mathrm{eff} + m({1+\alpha})/{2}$ to illustrate the fluxonium spectrum.

\begin{figure}[t!]
    \centering
    \includegraphics[width=1\linewidth]{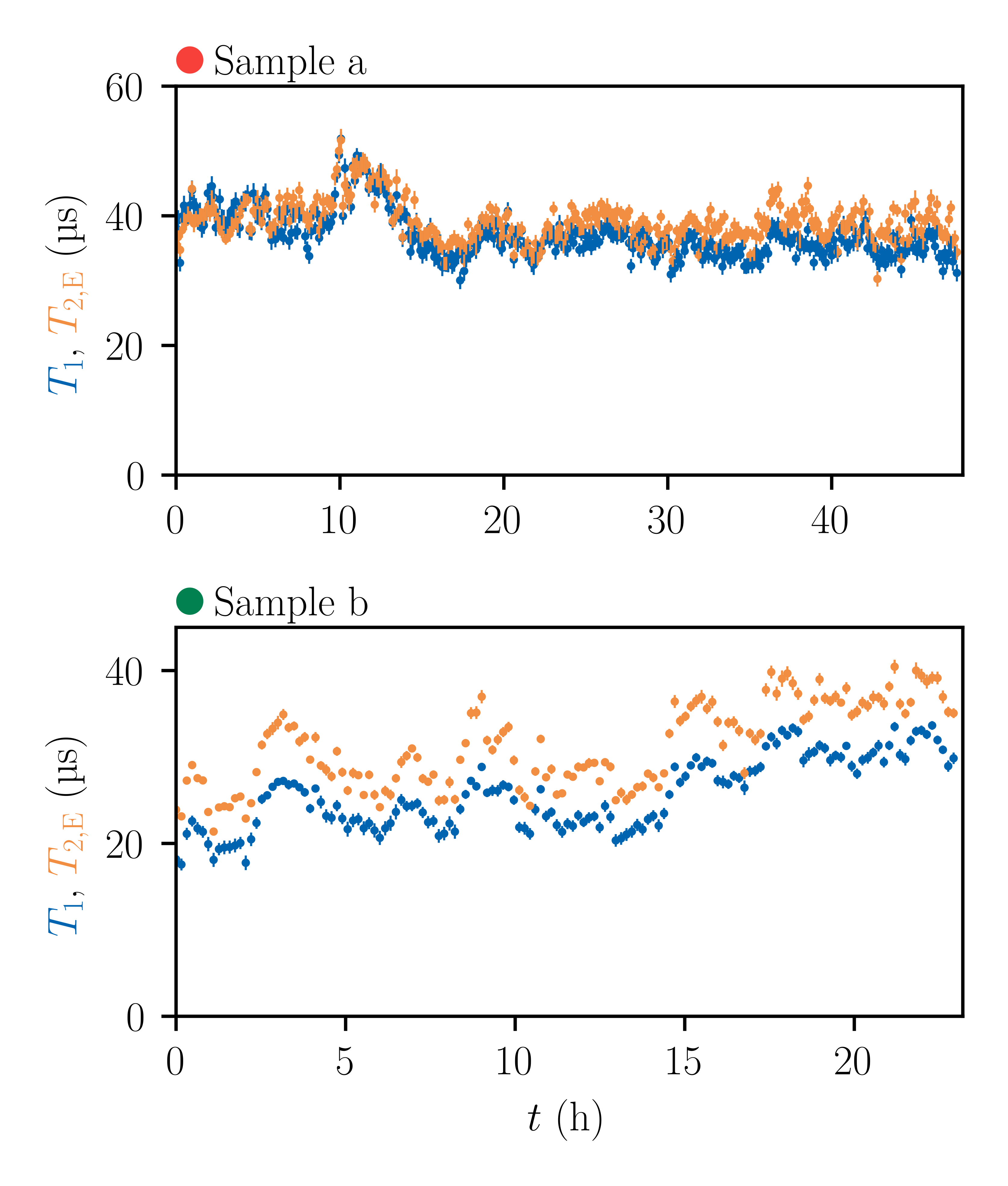}
    \caption{Interleaved $T_1$ and $T_\mathrm{2,E}$ coherence times at the sweet spot for samples \textit{a} and \textit{b} $\pi$-locked, over time. }
    \label{fig:coherence_over_time}
\end{figure}

\section{Asymmetry}
\label{subsec:asymmetry}

The asymmetry in gradiometric fluxoniums has both a geometric and an inductive contribution. The geometric contribution is linked with the exact path taken by the current in the loops and introduces a variation in $\Delta A = A'-A''$. The inductive contribution comes directly from the inductance of each loop and affects the value of $\alpha$.
Because $A_\mathrm{eff}={\Delta A}/{2} + {\alpha}/{2} \,(A'+A'')$, the effective area depends on both contributions.
For a better understanding, we plot in \cref{fig:DeltaL_DeltaA} the inductance difference (a) and the area difference (b) between each loop as a function of the asymmetry $({w'-w''})/{2}$ introduced in the main text.
Linear fits are as follows:
\begin{align}
    \Delta L &= (\SI{-120}{pH / \micro m}) \frac{w'-w''}{2} - \SI{8.5}{pH} \\
    \frac{\Delta A}{2} &= (\SI{92.7}{\micro m^2 / \micro m}) \frac{w'-w''}{2} + \SI{8.9}{\micro m^2} 
\end{align}
For reference, the linear fits in \Cref{fig:fig2} of the main text are:
\begin{align}
    \alpha &= (\SI{-4.4}{\percent / \micro m}) \frac{w'-w''}{2} - \SI{0.31}{\percent} \\
    A_\mathrm{eff} &= (\SI{23.6}{\micro m^2 / \micro m}) \frac{w'-w''}{2} + \SI{4.0}{\micro m^2} 
\end{align}

\section{Coherence}
\label{subsec:coherence_over_time}

After calibrating the position of the ground and excited states in the IQ plane with our single-shot readout, we calibrated a $\pi$-pulse and a $\pi/2$-pulse from Rabi oscillations when driving the qubit. 
$T_1$ was obtained by applying a $\pi$-pulse and measuring the relaxation of the qubit, which is then fitted to an exponential decay $e^{-t/T_1}$.
$T_\mathrm{2,E}$ was obtained from a Hahn-echo sequence consisting of two $\pi/2$-pulses with a $\pi$-pulse in between. At the sweet spot, $T_\mathrm{2,E}$ is fitted with only an exponential decay $e^{-\Gamma_\mathrm{2,E}^0 t}$. Away from the sweet spot, an additional gaussian decay $e^{-(\Gamma_\mathrm{2,E}^\Phi t)^2}$ is fitted on top of the exponential decay, and $T_\mathrm{2,E}$ is then calculated as:
\begin{equation}
    T_\mathrm{2,E} = \frac{\sqrt{(\Gamma_{2,E}^0)^2 + 4 (\Gamma_\mathrm{2,E}^\Phi)^2} - \Gamma_{2,E}^0}{2 (\Gamma_\mathrm{2,E}^\Phi)^2}
\end{equation}
$T_1$ and $T_\mathrm{2,E}$ were measured around zero field (\Cref{fig:fig3} of the main text), 
as well as at the sweet spot over 50 hours (\Crefadd{fig:coherence_over_time}{a}) and  23 hours (\Crefadd{fig:coherence_over_time}{b}). These measurements over time were performed interleaved, i.e. each few-minute-long measurement alternated between a $T_1$ sequence and a Hahn-echo sequence. 
Both samples show large fluctuations of their coherence times, which appear to be correlated between $T_1$ and $T_{2,\mathrm{E}}$.

\section{Flux escape}
\label{subsec:flux_escape}

\begin{figure}
    \centering
    \includegraphics[width=1\linewidth]{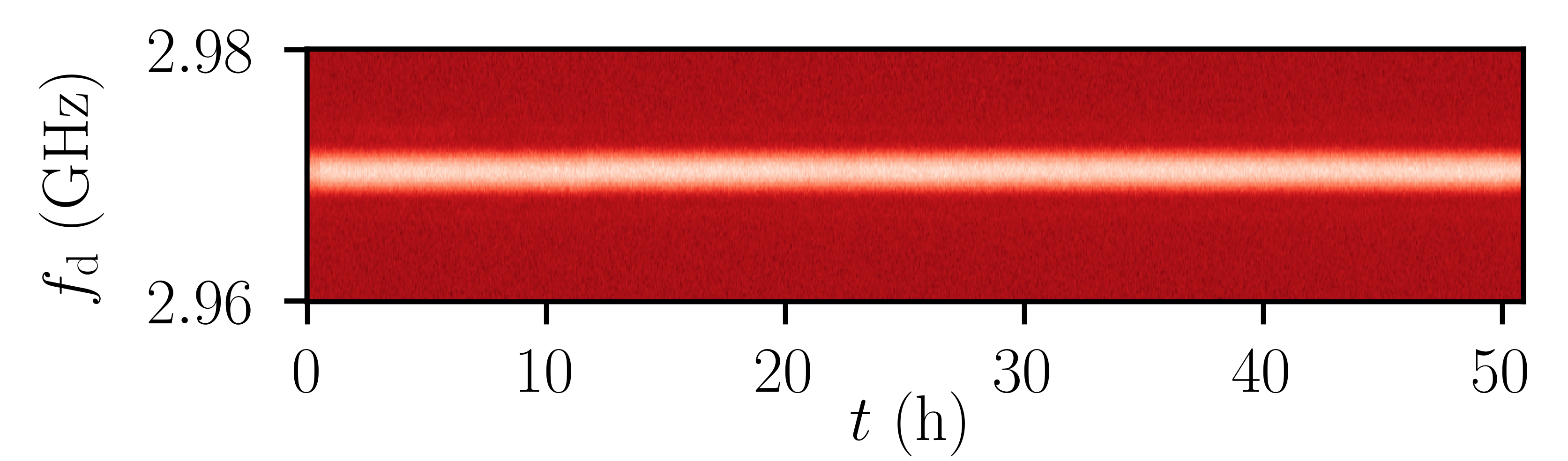}
    \caption{Response of the resonator for an additional sample, when the qubit is driven at frequency $f_\mathrm{d}$, in a standard two-tone spectroscopy. The qubit was biased to the sweet spot and measured over 50 hours. }
    \label{fig:two_tone_over_time}
\end{figure}

In order to confirm the absence of the flux escape phenomenon previously reported in Ref. \cite{gusenkova_operating_2022}, we repeated a two-tone spectroscopy over 2 days. The trace, shown in \cref{fig:two_tone_over_time}, does not show any flux escape over 50 hours. The measurement was performed on a sample fabricated in a different batch with similar evaporation parameters to the samples in the main text.

\section{Flux crosstalk}
\label{subsec:flux_crostalk}

\begin{figure}[htbp]
    \centering
    \includegraphics[width=1\linewidth]{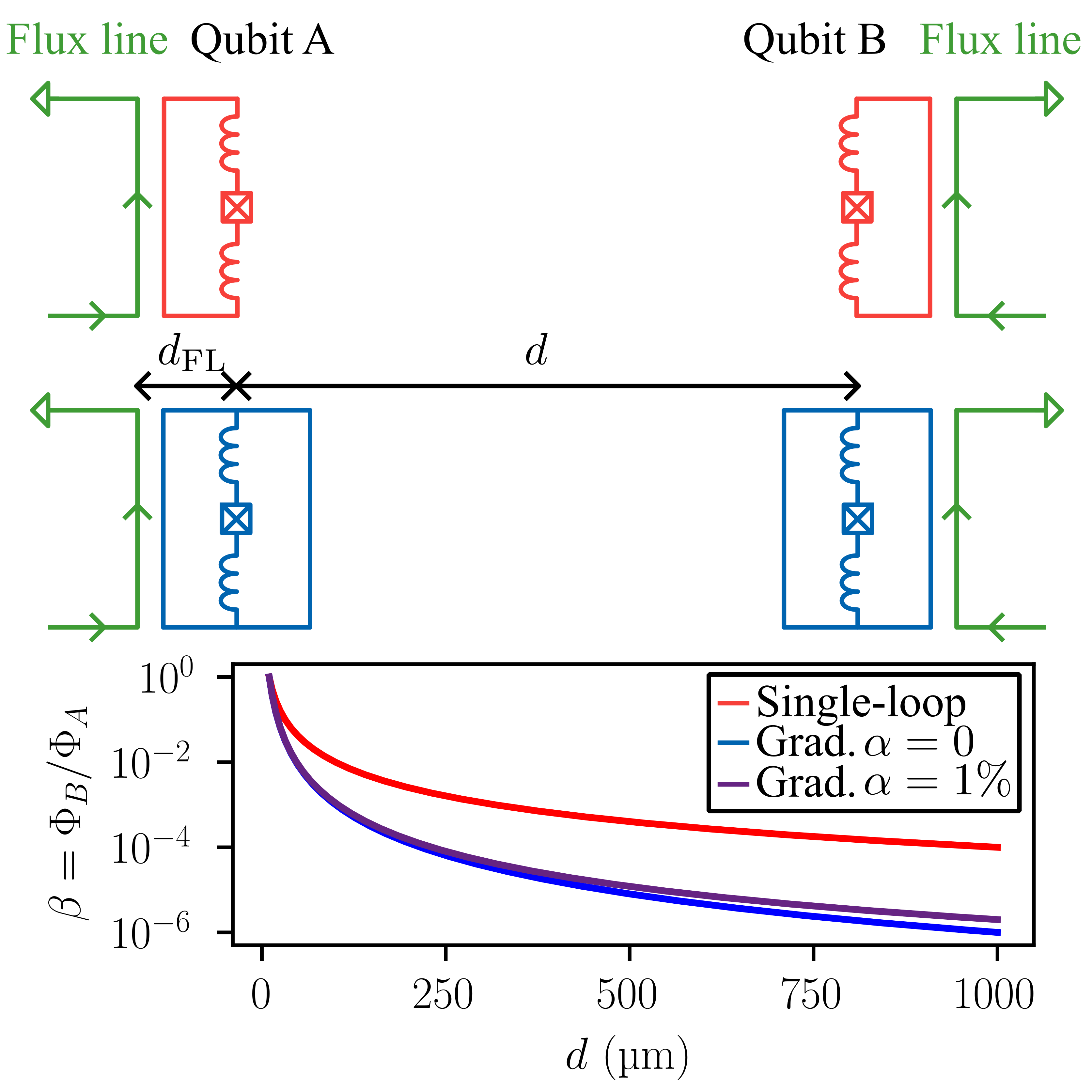}
    \caption{Generic diagram for a two-qubit chip. Top: two single-loop fluxonium qubits. Middle: two gradiometric fluxonium qubits. Bottom: Variation of the ratio $\beta=\Phi_B/\Phi_A$ for single-loop (red) and gradiometric with $\alpha=0$ (blue) and $\alpha=1\%$ (purple) fluxoniums.}
    \label{fig:flux_crosstalk}
\end{figure}

Let us take two fluxoniums separated by a distance $d$ (typically $d\approx\SI{500}{\micro m}$ in a quantum processor) and each of them at a distance $d_\mathrm{FL}$ from their flux line, as represented in \Cref{fig:flux_crosstalk}. 
For the regular one-loop fluxonium and the gradiometric fluxonium cases, we can calculate the flux $\Phi_\mathrm{A}$ seen by the qubit closest to the flux line and the flux $\Phi_\mathrm{B}$ seen by the qubit furthest from the flux line, for a given current $I$.
For one-loop fluxoniums, following the Biot-Savart law we get:
\begin{align*}
    \Phi_\mathrm{A} &\propto B(d_\mathrm{FL}) \propto \frac{1}{d_\mathrm{FL}^2} \\
    \Phi_\mathrm{B} &\propto B(d+d_\mathrm{FL}) \propto \frac{1}{(d+d_\mathrm{FL})^2}  
\end{align*}
Therefore, the flux crosstalk, characterized by the ratio $\beta = \Phi_\mathrm{B}/\Phi_\mathrm{A}$ is $\beta \approx (d_{\mathrm{FL}}/d)^2$, about \SI{4e-4}{} for $d=\SI{500}{\micro m}$ and $d_\mathrm{FL} = \SI{10}{\micro m}$. In contrast, for gradiometric fluxoniums what matters is the magnetic field gradient, and following the same argument we get:
\begin{align*}
    \Phi_\mathrm{A} &\propto \dv{B}{d} (d_\mathrm{FL})  \propto \frac{1}{d_\mathrm{FL}^3} \\
    \Phi_\mathrm{B} &\propto \dv{B}{d}(d+d_\mathrm{FL}) \propto \frac{1}{(d+d_\mathrm{FL})^3}  
\end{align*}
The flux crosstalk ratio is then $\beta \approx (d_{\mathrm{FL}}/d)^3 $, about $\SI{8e-6}{}$ for $d=\SI{500}{\micro m}$ and $d_\mathrm{FL} = \SI{10}{\micro m}$.
In summary, we gain about two orders of magnitude on the crosstalk because of the gradiometric nature of the devices.
Additionally, we gain three orders of magnitude on the current needed to bias each qubit to its sweet spot.

\section{Area asymmetry}
\label{subsec:area_asymmetry}

\begin{figure}[htbp]
    \centering
    \includegraphics[width=1\linewidth]{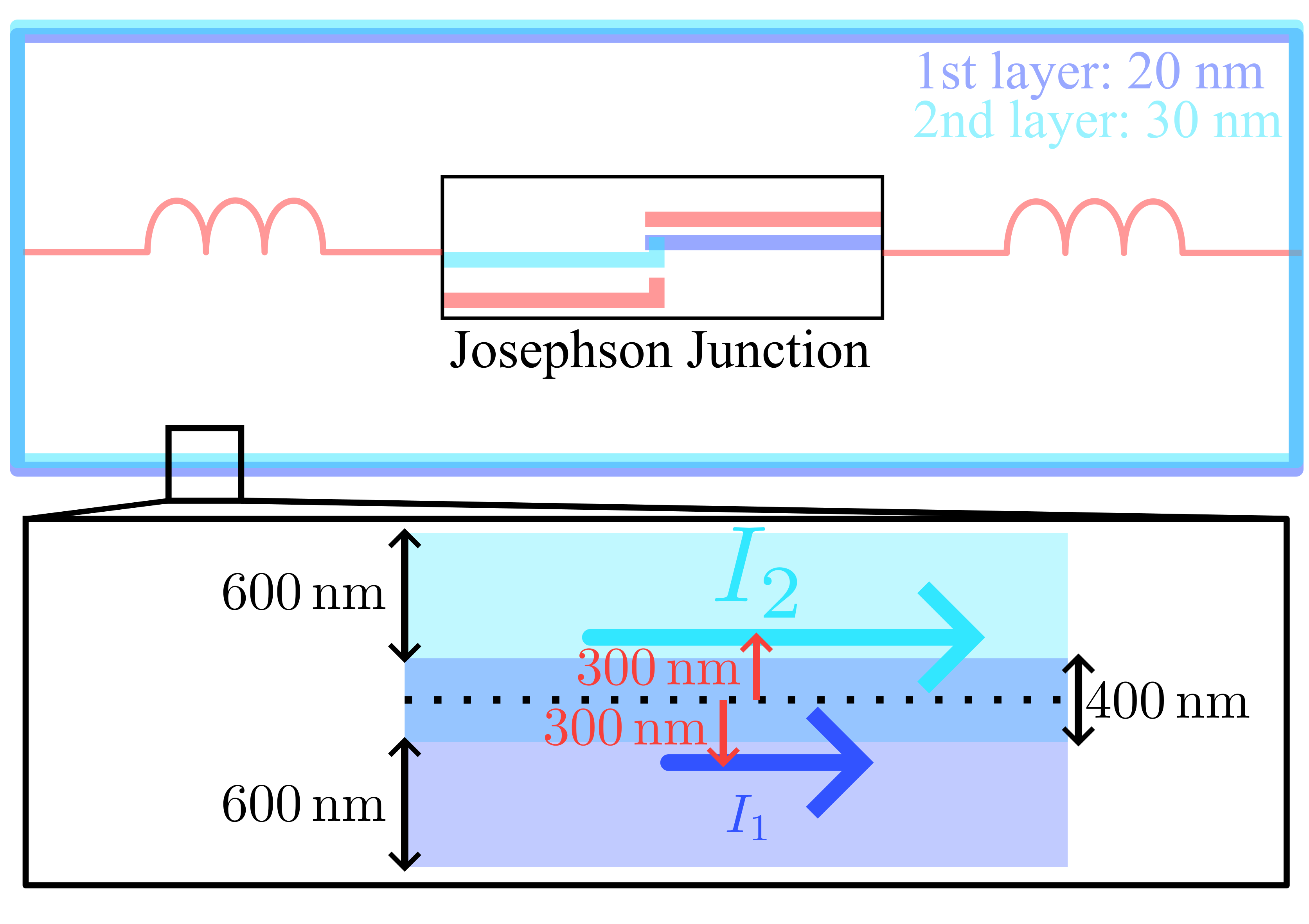}
    \caption{Sketch of the device with the aluminum layers in different colors to illustrate their different thicknesses. The arrows in the zoom-in illustrate the positions of the persistent current center lines in each  layer. The length of each arrow gives the magnitude of the current.}
    \label{fig:area_asymmetry}
\end{figure}

In order to fabricate the SIS Josephson junction in the middle of the qubit, we evaporate two layers of aluminum, with tilt angles $\SI{-31}{\degree}$ and $\SI{31}{\degree}$. This shadow evaporation creates a stack of 20-nm-thick and 30-nm-thick aluminum layers, respectively, creating two paths for the supercurrent with different kinetic inductances, as illustrated in Fig. \ref{fig:area_asymmetry}. We estimate, from the empirical scaling of Ref. \cite{lopez-nunez_magnetic_2023}, the sheet inductance of each layer to be $\SI{2.1}{pH}/\square$ and $\SI{0.9}{pH}/\square$, respectively. The current is thus twice as large in the 30-nm-thick layer as in the 20-nm-thick layer: $I_2/I_1=2$. The aluminum layers are $\SI{600}{nm}$ apart, with $\SI{400}{nm}$ overlap, i.e. the middle of the thicker layer is $\delta=\SI{300}{nm}$ away from the middle of the bilayer stack. Correspondingly, the effective offset for the persistent current path is $\delta \, (1-I_1/I_2)/(1+I_1/I_2)=\SI{100}{nm}$, close to the value we find experimentally.

%


\end{document}